\documentstyle[12pt]{article}

\begin{document}
\title{Vacuum Polarization at Finite Temperature around a Magnetic Flux
Cosmic String}
\author{M. E. X. Guimar\~aes \thanks{electronic address:
marg@ccr.jussieu.fr} \\
\mbox{\small Laboratoire de Gravitation et Cosmologie Relativistes,
CNRS/URA 769} \\
\mbox{\small Universit\'e Pierre et Marie Curie, Tour 22/12, B. C. 142} \\
\mbox{\small 4, Place Jussieu, 75252, Paris cedex 05, France}}
\maketitle
\begin{abstract}
We consider a general situation where a charged massive scalar field
$\phi(x)$ at finite
temperature interacts with a
magnetic flux cosmic string. We determine a general expression
for the Euclidean thermal Green's function of the massive scalar field and a
handy expression for a massless scalar field. With this result, we
evaluate the thermal average $<\phi^{(2)}(x)>$ and the thermal
average of the energy-momentum tensor of a nonconformal massless
scalar field.

{\em Classification PACS: 03.70 and 98.80C}
\end{abstract}

\section{Introduction}

In the General Relativity framework, a straight, infinite, static string,
lying on
the z-axis, is described by a static metric with cylindrical symmetry

\begin{equation}
ds^2= -dt^2+dz^2+d\rho^2+B^2\rho^2d\varphi^2
\end{equation}
where $\rho \geq 0$ and $0 \leq \varphi < 2\pi$ and
the constant $B$ is related to the linear mass density $\mu$ as $B=1-4\mu$
\footnote{We use here the system of units $G=\hbar = c=1$.}. Spacetime (1)
is locally but not globally flat \cite{VI}. The presence of the string leads
to an azymuthal deficit angle $8\pi\mu$ and, as a result,
this spacetime has a conical singularity.

One of the most remarkable features of spacetime (1) is the fact that
fields are
sensitive to its global structure and interesting effects can occur
\cite{SMI}. For
instance, in the context of Quantum Field Theory in curved spacetimes, the
vacuum state of a quantum field is polarized due solely to the presence of
the string in this geometry. This problem has been studied in many papers
for
various massless fields, including the case where the cosmic string
carries an internal magnetic flux (for a review, see Dowker \cite{DO}).

Recently, Guimar\~aes and Linet \cite{GL} have considered a more general
situation where a charged massive scalar field interacts with a cosmic string
carrying an internal magnetic flux $\Phi$. They have found the Euclidean
Green's function
$G_{E}^{(\Phi)}(x, x_{0}, m)$ of the massive scalar field in a local form
in which $G_E^{(\Phi)}(x, x_{0}, m)$ is
the sum of the usual Euclidean Green's function and a regular term
$G_E^{* (\Phi)}(x, x_{0}, m)$, for $x$ close enough to $x_0$. This form
is very convenient
because the renormalization is straightforward and consists of removing the
usual Euclidean Green's function from the expression of
$G_E^{(\Phi)}(x, x_{0}, m)$.
Thus, the calculus of the vacuum expectation values of the
energy-momentum tensor
consists in taking the coincidence limit $x=x_{0}$ of the regular term and
its derivatives.

The phenomenom of vacuum polarization of a quantum field by a cosmic string
carrying an internal magnetic flux
can be understood as a
realization in Cosmology of the current Aharonov-Bohm effect \cite{AB}.
Indeed, a quantum field placed in the exterior region of the string acquires
an additional phase shift proportional to the magnetic flux even
thought there is no magnetic field outside the string \cite{SE}.
Our purpose in this paper is to analyse the vacuum polarization of a
nonconformal, charged massive scalar field at finite temperature
by a magnetic flux cosmic string. Recently, Frolov et al. \cite{FR}
considered
the vacuum polarization of a nonconformal, massless
scalar field at finite temperature on
a cone. Our work differs from theirs in two important aspects. First, we
consider a cosmic string which carries an internal magnetic flux.
Second, we
restrict our calculations to values of constant $B$ such that $B>1/2$,
while
their calculations is for arbitrary values of $B$. However, it must be
remarked that
setting $B>1/2$ represents no physical restriction because strings of
cosmological interest are of order $\mu \sim 10^{-6}$.

To study the vacuum polarization phenomenom, we need first to compute the
thermal Green's function of the scalar field. Since it is more convenient to
work in the Euclidean approach to quantum theory, we will, in practice,
compute the
Euclidean thermal Green's function of the scalar field. This represents no
complication because, spacetime (1) being globally static, the Euclidean
metric is
obtained straightforwardly by a Wick rotation ($t= -i\tau$) in the
coordinate $t$ of metric (1).
The method used here is the same
as in Linet's paper \cite{LI} which works in the framework of
the Schwinger-DeWitt formalism \cite{SDW}. Namely, the Euclidean thermal
Green's function $G_{ET}^{(\Phi)}(x, x_{0},m)$
is calculated from its corresponding Euclidean thermal heat kernel
$K_{ET}^{(\Phi)}(x, x_{0},s)$ by an integral on the variable $s$. Moreover,
since the metric we are dealing
with is ultrastatic (static and $g_{00}=1$),
the Euclidean thermal heat kernel $K_{ET}^{(\Phi)}(x, x_{0}, s)$ can be
derived from
the Euclidean zero-temperature heat kernel $K_{E}^{(\Phi)}(x, x_{0}, s)$
\cite{PA}. Thus, our task is, first of all, to determine
$K_{E}^{(\Phi)}(x, x_{0}, s)$.
Since in the paper \cite{GL}, the Euclidean zero-temperature
Green's function
$G_{E}(x, x_{0}, m)$ was already determined, we can easily find
$K_{E}(x, x_{0}, s)$
from it.

This paper is organized as follows. In section II, we briefly review
the Euclidean Green's function at zero-temperature. Then, we derive
its corresponding Euclidean heat kernel.
In section III, we determine the Euclidean thermal heat kernel and its
corresponding Green's function.
In section IV, by applying the results obtained in section
III we give the thermal average $<\phi^{2}(x)>$ in a general
form. Then, we give both zero and high-temperatures limits of this
general expression.
In section V, we determine the thermal average of the energy-momentum
tensor of a nonconformal massless scalar field. We provide the
general expression and then make the zero and high-temperatures limits.
We compare our results in the case where the flux vanishes with
the ones of Frolov et al. \cite{FR}
when $B>1/2$. In section VI, we add some concluding remarks.

\section{Preliminaries}

We consider the four-dimensional Riemannian metric described by

\begin{equation}
\label{euc}
ds^2=d\tau^2+dz^2+d\rho^2+B^2\rho^2d\varphi^2,
\end{equation}
in a coordinate system $(\tau , z, \rho , \varphi)$, with $\rho \geq 0$ and
$0 \leq \varphi < 2\pi$, which is obtained from (1) by means of a
Wick rotation $(t= -i\tau)$ in the coordinate $t$. As we already
mentioned in the
introduction, metric (\ref{euc}) has a global conical structure.
Moreover, we
will consider that the cosmic string, which is the source of geometry (2),
carries an internal magnetic flux $\Phi$.

The Euclidean Green's function of a charged
massive scalar field $G_{E}^{(
\gamma)}
(x, x_{0}, m)$ obeys the covariant Laplace equation in the space
described by metric (2)

\begin{equation}
[\frac{\partial^{2}}{\partial{\tau^{2}}}+\frac{\partial^2}{\partial{z^2}}+
\frac{\partial^2}{\partial{\rho^2}}+\frac{1}{\rho}
\frac{\partial}{\partial{\rho}}
+\frac{1}{B^2\rho^2}\frac{\partial^2}{\partial{\varphi^2}}-m^2]
G_{E}^{(\gamma)}(x, x_{0}, m)
= - \frac{1}{B}\delta^{(4)}(x, x_{0})
\end{equation}
and satisfies the
following boundary conditions

\[
G_{E}^{(\gamma)}(\tau , z, \rho , \varphi + 2\pi) = e^{2i\pi\gamma}
G_{E}^{(\gamma)}
(\tau , z, \rho , \varphi )
\]
\begin{equation}
\frac{\partial}{\partial{\varphi}}
G_{E}^{(\gamma)}(\tau, z, \rho , \varphi + 2\pi) =
e^{2i\pi\gamma}\frac{\partial}{\partial{\varphi}}
G_{E}^{(\gamma)}(\tau , z,
\rho , \varphi ),
\end{equation}
where $\gamma$ is the fractional part of $\Phi /\Phi_{0}$, $\Phi_{0}$ being
the flux quantum $2\pi /e$, and lies in the interval $0 \leq \gamma < 1$.
We shall remark that the case $\gamma = 0$
represents the situation where the flux is vanishing and the case
$\gamma = 1/2$
describes a twisted scalar field around the axis $\rho =0$ [2].
We also require that $G_{E}^{(\gamma)}(x, x_{0}, m)$ must
vanish when $x$ and $x_{0}$ are infinitely separated.

An Euclidean Green's function which is solution to equation (3) and
that satisfies
requirements (4) was already obtained in the paper of ref. [4].
When $B>1/2$,
it is presented
in a local form valid in the domain $\frac{\pi}{B}-2\pi <
\varphi -\varphi_0 <2\pi -\frac{\pi}{B}$ as a sum of the usual
Green's function in Euclidean space and a regular term, which is the
net effect
of the conical structure of (2). That is

\begin{equation}
G_{E}^{(\gamma)}(x, x_{0}, m) = \frac{mK_{1}[mr_{4}]}{4\pi^2r_{4}} +
G_{E}^{*(\gamma)}(x, x_{0}, m),
\end{equation}
where the regular term is
\[
G_{E}^{*(\gamma)}(x, x_0, m) = \frac{m}{8\pi^3 B}\int_{0}^{\infty}
\frac{K_{1}[ mR_{4}(u)]}{R_{4}(u)}F_{B}^{(\gamma)}(u,\psi) du ,
\]
and $r_4$, $R_{4}(u)$ and $F_{B}^{(\gamma)}(u, \psi)$ are respectively
\[
r_4 = [(\tau -\tau_0)^2+(z-z_0)^2+\rho^2+\rho_0^2-2\rho\rho_0
\cos B(\varphi - \varphi_0)]^{1/2} ,
\]
\begin{equation}
R_{4}(u)=[(\tau -\tau_0)^2+(z-z_0)^2+\rho^2+\rho_0^2+
2\rho\rho_0\cosh u]^{1/2} ,
\end{equation}
\begin{eqnarray*}
F_{B}^{(\gamma)}(u, \psi) & = & i \frac{e^{i(\psi + \pi/B)\gamma}
\cosh[u(1-\gamma)
/B] - e^{-i(\psi +\pi/B)(1-\gamma)}\cosh[u\gamma/B]}{\cosh u/B -
\cos(\psi +\pi/B)}
 \\
& & -i\frac{e^{i(\psi -\pi/B)\gamma}\cosh [u(1-\gamma)/B]-
e^{-i(\psi -\pi/B)(1-\gamma)} \mbox{}
\cosh[u\gamma/B] }{ \cosh u/B -\cos(\psi -\pi/B)},
\end{eqnarray*}
with $ \psi \equiv \varphi -\varphi_0 $.

Thus, at this point, we are able to apply the Schwinger-DeWitt
formalism [9]. Namely,

\begin{equation}
G_{E}^{(\gamma)}(x, x_0, m)=\int_{0}^{\infty}
K_{E}^{(\gamma)}(x, x_0 ,s) ds .
\end{equation}
We remind that the heat kernel obeys the differential equation

\[
(\frac{\partial}{\partial {s}} - \Box +m^2)K^{(\gamma)}_{E} = 0,
\]
for $s>0$ and satisfies the initial condition
\[
K_{E}^{(\gamma)}(x, x_0, 0)= \frac{1}{B} \delta^{(4)}(x, x_0).
\]
Taking into account the local form (5) of $G_{E}^{(\gamma)}(x, x_0 ,m)$,
we can see that the corresponding heat kernel $K_{E}^{(\gamma)}
(x, x_0 , s)$ will be also represented as a sum of the usual heat kernel in
Euclidean space and a regular term. Besides, using the Laplace
inverse transformation

\[
\frac{\sqrt{p} K_{1}[d\sqrt{p}]}{d}=\int_{0}^{\infty}\frac{1}{4s^2}
e^{-d^2/4s} e^{-ps}ds ,
\]
and inserting in expression (7), with $G^{(\gamma)}_{E}(x, x_0 ; m)$
given by (5) we deduce

\begin{equation}
K_{E}^{(\gamma)}(x, x_0 ; s)= \frac{1}{16\pi^2 s^2}e^{-r_{4}^{2}/4s}
e^{-m^2 s} + K_{E}^{*(\gamma)}(x, x_0 ; s) ,
\end{equation}
where the regular term is
\[
K_{E}^{*(\gamma)}(x, x_0 ; s) = \frac{1}{32\pi^3 Bs^2}e^{-m^2 s}
\int^{\infty}_{0}
e^{-R_{4}^{2}(u)/4s}F_{B}^{(\gamma)}(u,\psi) du .
\]
for $B>1/2$ and for $x$ close enough to $x_0$.

Thus, we have determined an integral expression (8) the Euclidean
(zero-temperature)
heat kernel corresponding to the Euclidean (zero-temperature)
Green's function
(5) of a charged, massive scalar field in the Riemannian metric (2).

\section{The Euclidean Thermal Heat Kernel and its Corresponding
Green's Function}

Just as a remainder of the paper, we recall that the
thermal Green's function of a charged scalar field
$G_{ET}^{(\gamma)}(x, x_0 ;m)$
is a solution to equation (3), satisfies the boundary conditions (4),
vanishes for spatial points $x^{i}$ and $x^{i}_0$ infinitely
separated and morevover,
it is periodic in the coordinate $\tau$ with a period given by
$\beta = 1/kT$, $k$ and
$T$ being respectively the Boltzmann constant and the temperature.
To obtain
$G_{ET}^{(\gamma)}(x, x_0 , m)$ we will first derive the Euclidean
thermal heat kernel
$K_{ET}^{(\gamma)}(x, x_0 ; s)$, which  can be
determined straightforwardly from its zero-temperature limit
$K_{E}^{(\gamma)}(x, x_0 ; s)$.
Indeed, as shown in papers \cite{PA}, for ultrastatic
metrics such as (2), $K_{ET}^{(\gamma)}(x, x_0 ; s)$ can
be factorized in the following way

\begin{equation}
K_{ET}^{(\gamma)}(x, x_0 ; s)= \Theta_{3}(i\frac{\beta (\tau -\tau_{0})}{4s}
\mid i\frac{\beta^{2}}{4\pi s})K_{E}^{(\gamma)}(x, x_0 ;s) ,
\end{equation}
where $\Theta_{3}$ is the Theta function \cite{GR} defined as
\[
\Theta_{3}(z \mid t)=\sum_{n=-\infty}^{\infty} e^{i\pi n^{2} t+2inz} .
\]

Substituting (8) into (9), we have

\begin{equation}
K_{ET}^{(\gamma)}(x, x_0 ;s) = \frac{1}{16\pi^2 s^2}
\Theta_{3}(i\frac{\beta(\tau -\tau_0)}{4s}\mid
i\frac{\beta^2}{4\pi s})
e^{-r_{4}^2/4s} e^{-m^{2}s} +
K_{ET}^{*(\gamma)}(x, x_0 ;s) ,
\end{equation}
with
\[
K_{ET}^{*(\gamma)}(x, x_0 ;s) =\frac{1}{32\pi^3 Bs^2}
\Theta_{3}(i\frac{\beta(\tau -\tau_0)}{4s}\mid
i\frac{\beta^2}{4\pi s})
e^{-m^2 s}\int^{\infty}_{0}e^{-R_{4}^{2}(u)/4s} F_{B}^{(\gamma)}(u,\psi) du.
\]
Applying formula (7) we finally find
the expression of the Euclidean thermal Green's function for the massive
scalar field

\begin{equation}
G_{ET}^{(\gamma)}(x, x_0 ;m) = \frac{1}{16\pi^2}\int^{\infty}_{0}
\Theta_{3}(i\frac{\beta(\tau -\tau_0)}{4s}\mid
i\frac{\beta^2}{4\pi s})
e^{-r_{4}^{2}/4s} \frac{e^{-m^2 s}}{s^2} ds
+ G_{ET}^{*(\gamma)}(x, x_0 ; m) ,
\end{equation}
where the regular term is
\[
 G_{ET}^{*(\gamma)}(x, x_0 ; m) = \frac{1}{32\pi^3 B} \int_{0}^{\infty}
 \int_{0}^{\infty}
\Theta_{3}(i\frac{\beta(\tau -\tau_0)}{4s}\mid
i\frac{\beta^2}{4\pi s})
e^{-R_{4}^{2}(u)/4s}\frac{e^{-m^2 s}}{s^2}F_{B}^{(\gamma)}(u,\psi) ds du .
\]

Now, expression (11) cannot be integrated. A handy form will be provided in
the case where the mass of the scalar field tends to zero.
So that, we set
$m=0$ in expression (11) and we obtain

\begin{equation}
D_{ET}^{(\gamma)}(x, x_0)=\frac{1}{16\pi^2}
\int_{0}^{\infty}
\frac{1}{s^2}
e^{-r_{4}^{2}/4s}
\Theta_{3}(i\frac{\beta(\tau -\tau_0)}{4s}\mid
i\frac{\beta^2}{4\pi s}) ds +
D_{ET}^{*(\gamma)}(x, x_0) ,
\end{equation}
with the regular term
\[
D_{ET}^{*(\gamma)}(x, x_0) = \frac{1}{32\pi^3 B}\int^{\infty}_{0}
\int_{0}^{\infty}
\frac{1}{s^2} e^{-R_{4}^{2}(u)/4s}
\Theta_{3}(i\frac{\beta(\tau -\tau_0)}{4s}\mid
i\frac{\beta^2}{4\pi s}) F_{B}^{(\gamma)}(u,\psi) ds du .
\]

Following Linet's procedure in paper \cite{LI}, we change the
variable $s$ for $t$ in the way
\[
t= \frac{\beta^{2}}{4\pi^2} \frac{1}{s} ,
\]
and substitute in expression (12). We obtain

\begin{eqnarray}
& &  D_{ET}^{(\gamma)}(x, x_0) =  \frac{1}{4\beta^{2}}
\int_{0}^{\infty} \Theta_{3}(i\pi\kappa t\mid i\pi t)
e^{-(d^{2}+\kappa^{2})t} dt \nonumber \\
& & +  \frac{1}{8 \pi B \beta^{2}}
\int^{\infty}_{0} \int_{0}^{\infty} \Theta_{3}(i\pi\kappa t\mid i\pi t)
e^{-(D^{2}(u) + \kappa^{2})t} F^{(\gamma)}_{B}(u, \psi) du dt ,
\end{eqnarray}
where we define the functions $d, D(u)$ and $ \kappa $ respectively as
\[
d = \frac{\pi}{B} [(z-z_0)^{2}+\rho^{2}+\rho_{0}^{2}-2\rho\rho_{0}
\cos B(\varphi - \varphi_{0})]^{1/2} ,
\]
\begin{equation}
D(u) = \frac{\pi}{B} [(z-z_{0})^{2}+\rho^{2}+\rho_{0}^{2}+2\rho\rho_{0}
\cosh u]^{1/2} ,
\end{equation}
\[
\kappa = \frac{\pi}{B}(\tau -\tau_{0}) .
\]

Let's us turn our attention to the integral
\[
I(a,b) = \int_{0}^{\infty}\Theta_{3}(i\pi b t\mid i\pi t)
e^{-(a^{2}+b^{2})t} dt,
\]
which appears in both terms of expression (13). This integral
can be performed and gives
\begin{equation}
I(a,b) = \frac{1}{2a}[\coth (a+ib) + \coth (a -ib)] .
\end{equation}
Substituting this result in (13), we get the final form of the
Euclidean thermal Green's function for a massless scalar field

\begin{eqnarray}
& & D_{ET}^{(\gamma)}(x, x_0)  =  \frac{1}{4\pi\beta d}
\frac{\sinh (2\pi / \beta) d}
{[\cosh (2\pi / \beta) d - \cos (2\pi / \beta)(\tau - \tau_0) ]}
\nonumber \\
& & + \frac{1}{8\pi^2 B \beta}\int_{0}^{\infty}
\frac{\sinh [ 2\pi /\beta D(u)] F_{B}^{(\gamma)}
(u, \psi)}{D(u) [\cosh 2\pi /\beta D(u)
- \cos 2\pi /\beta (\tau -\tau_0)]}  du ,
\end{eqnarray}
for $B> 1/2$.

We must point out that, in the limite where the temperature goes to zero,
the above result tends to the Euclidean Green's function $D_{E}^{(\gamma)}
(x, x_{0})$, already found in the paper of reference \cite{GL}.

\section{The Thermal Average $< \phi^{2} (x) >_{\beta}$}

As an application of the results obtained in the previous section, we can
evaluate the thermal average  $< \phi^{2} (x) >_{\beta}$.
The procedure consists of removing the usual (zero-temperature)
Green's function in the coincident limit $x=x_0$:

\begin{equation}
< \phi^{2} (x) >_{\beta} = \lim_{x \rightarrow x_0}
[ D_{ET}^{(\gamma)}(x, x_0) - 1/ 4\pi^{2}r_{4}^{2}] .
\end{equation}
{}From (16) we obtain the general expression for the thermal average

\begin{equation}
< \phi^{2} (x) >_{\beta} = \frac{1}{12 \beta^{2}} +
\frac{1}{16\pi^2 B\beta \rho}\int_{0}^{\infty}
\frac{\coth [\frac{2\pi}{\beta}
\rho\cosh u/2]}{\cosh u/2} F_{B}^{(\gamma)}(u, 0) du ,
\end{equation}
where
\[
F_{B}^{(\gamma)}(u, 0) = -2 \frac{\sin[\pi\gamma /B]\cosh [u(1-\gamma)/B]+
\sin[\pi (1-\gamma)/B]\cosh [u\gamma/B]}{\cosh u/B -\cos \pi/B} .
\]
We can recognize in expression (18) the usual thermal average in Euclidean
space and the second term which is particular from the conical structure.

Unfortunately, we cannot integrate expression (18). Therefore, we will give
some asymptotic limits of it. First, we consider
the zero-temperature limit ($\beta \rightarrow \infty$), we have

\begin{equation}
< \phi^{2} (x) >_{\beta\rightarrow \infty} \sim \frac{\omega_{2}(\gamma)}{2
\rho^2},
\end{equation}
where we define some quantities which will be useful later
\[
\omega_{2}(\gamma) \equiv \frac{1}{16\pi^3 B}\int_{0}^{\infty}
\frac{F_{B}^{(\gamma)}(u, 0)}{\cosh^{2} u/2} du ,
\]
\[
\omega_{4}(\gamma) \equiv \frac{1}{32\pi^{3}B}\int_{0}^{\infty}\frac{F_{B}^
{(\gamma)}(u,0)}{\cosh^{4}u/2}du .
\]
We remind that the constants $\omega_2(\gamma)$ and $\omega_4(\gamma)$ have
already appeared in ref. \cite{GL}
but their explicit expression have been found previoulsy by Dowker \cite{DW}
\[
\omega_{2}(\gamma)=-\frac{1}{8\pi^{2}}{\frac{1}{3}-\frac{1}{2B^{2}}[
4(\gamma -\frac{1}{2})^{2}-\frac{1}{3}]} ,
\]
\begin{eqnarray*}
\omega_{4}(\gamma) & = & -\frac{1}{720\pi^{2}} \{11-\frac{15}{B^{2}}
[ 4 ( \gamma -\frac{1}{2})^{2}-\frac{1}{3}] \nonumber \\
& & +\frac{15}{8B^{4}} [16(\gamma -\frac{1}{2})^{4}-
8( \gamma -\frac{1}{2})^{2}
+\frac{7}{15}] \} ,
\end{eqnarray*}
for $B>1/2$. If we set $\gamma = 0$ and $\gamma = 1/2$ in
expression (19) we get the same results
as Smith \cite{SMI}.

We consider now the high-temperature limit
($\beta \rightarrow 0$).
We have

\begin{equation}
< \phi^{2} (x) >_{\beta\rightarrow 0} \sim \frac{1}{12\beta^2} +
\frac{M^{(\gamma)}}{\beta\rho}  ,
\end{equation}
where we define the following constants $M^{(\gamma)}$ and
$N^{(\gamma)}$, which
will be useful in further applications

\[
M^{(\gamma)} \equiv \frac{1}{16\pi^{2}B}\int_{0}^{\infty}
\frac{F_{B}^{(\gamma)}(u, 0)}{\cosh u/2} du ,
\]
and
\begin{equation}
N^{(\gamma)} \equiv \frac{1}{32\pi^2 B}\int_{0}^{\infty}
\frac{F_{B}^{(\gamma)}(u,0)}{\cosh^{3} u/2} du .
\end{equation}
We obtain an explicit expression of (20) for first
order in the angle deficit. For $\gamma$ defined in the
interval $0< \gamma <1$ and in the limit where $B \rightarrow 1$, we get

\begin{eqnarray}
< \phi^{2} (x) >_{\beta\rightarrow 0} & \sim & \frac{1}{12\beta^{2}}
+ \frac{(1-B)}{8\pi\beta\rho}[\gamma\Gamma(\frac{5}{2}-\gamma)\Gamma(\frac
{1}{2}+\gamma)
\nonumber \\
& & + (1-\gamma)\Gamma(\frac{3}{2}+\gamma)\Gamma(\frac{3}{2}-\gamma)] ,
\end{eqnarray}
$\Gamma(z)$ being the Gamma function \cite{GR}.
In the case where $\gamma \rightarrow 0$, we get the same result as Linet
\cite{LI}, the constants $M^{(\gamma)}$ and $N^{(\gamma)}$
reducing to $M$ and $N$ defined in the same paper. When $\gamma = 1/2$, we
find the same as Rogatko \cite{RO}.

\section{The Thermal Average $< T_{\nu}^{\mu}(x)>_{\beta}$}

The renormalized
energy-momentum tensor of a nonconformal, charged, massless scalar
field can be
derived from its Green's functions and derivatives in the following way

\begin{equation}
< T_{\nu}^{\mu}(x)>_{\beta} = \lim_{x\rightarrow x_0} D_{\nu}^{\mu}
[D_{ET}^{(\gamma)}(x,x_0)-1/4\pi^{2}r_4^{2}] ,
\end{equation}
where $D_{\nu}^{\mu}$ is a differential operator given by

\[
D^{\mu}_{\nu} = 2[(1-2\xi)\nabla_{\nu}\nabla^{\mu_{0}}-
2\xi\nabla^{\mu}_{\nu}
+(2\xi -1/2)\delta_{\nu}^{\mu}\nabla_{\alpha}\nabla^{\alpha_{0}}] ,
\]
$\nabla_{\nu}(\nabla^{\nu_{0}})$ corresponding to the covariant
derivative with
respect to the coordinate $x(x_0)$ and $\xi$ being the coupling parameter.

As in the case of the Green's functions and thermal average $< \phi^{2}(x)>
_{\beta}$,
$<T_{\nu}^{\mu}(x)>_{\beta}$ will be represented as a sum of the
thermal radiation in flat spacetime
and a term which corresponds to the polarization effect due to the conical
singularity. Thus, we have

\begin{equation}
< T_{\nu}^{\mu}(x)>_{\beta}  = \frac{\pi^{2}}{45\beta^{4}}diag(-3,1,1,1) +
< T_{\nu}^{* \mu}(x)>_{\beta}
\end{equation}

The second term of (24) is

\begin{equation}
< T_{\nu}^{* \mu}(x)>_{\beta} = D^{\mu}_{\nu}
D^{*(\gamma)}_{ET}(x,x) .
\end{equation}

The calculus are combersome but straightforward. We present  directly our
results. From (25) we get

\begin{equation}
 < T_{\tau}^{* \tau}(x)>_{\beta}  =  I_1 +
(2\xi -1/2)[2 I_2 +  I_3 + I_4]
\end{equation}
\begin{equation}
< T_{z}^{* z}(x)>_{\beta} = I_5 +I_6 +
(2\xi -1/2)[2 I_2 +  I_3 + I_4]
\end{equation}
\begin{equation}
< T_{\rho}^{* \rho}(x)>_{\beta} = I_5 + I_6 -4\xi [I_3 + I_4]
\end{equation}
\begin{equation}
< T_{\varphi}^{* \varphi}(x)>_{\beta} = -2 [ I_5 +I_6 + \frac{1}{2}I_1]
+ 2\xi [ I_2 + I_3 + I_4]
\end{equation}
where
\[
I_1  \equiv  \frac{1}{4B\beta^{3}\rho}\int_{0}^{\infty}
\frac{\coth[\frac{2\pi}
{\beta}\rho\cosh u/2]}{\sinh^{2}[\frac{2\pi}{\beta}\rho\cosh u/2] \cosh u/2}
F_{B}^{(\gamma)}(u,0) du  \]
\[ I_2  \equiv \frac{1}{4B\beta^{3}\rho}\int_{0}^{\infty}
\frac{\coth[\frac{2\pi}{\beta}\rho\cosh u/2]\cosh u/2}
{\sinh^{2}[\frac{2\pi}{\beta}\rho\cosh u/2]}
F_{B}^{(\gamma)}(u,0) du \]
\[ I_3 \equiv \frac{1}{8\pi B \beta^{2}\rho^{2}}
\int_{0}^{\infty}\frac{1}{\sinh^{2}
[\frac{2\pi}{\beta}\rho\cosh u/2]}F_{B}^{(\gamma)}(u,0) du \]
\[
I_4 \equiv \frac{1}{16\pi^{2} B\beta\rho^{3}}\int^{\infty}_{0}
\frac{ \coth[ \frac{2\pi}{\beta}\rho\cosh u/2]}{\cosh u/2}
F_{B}^{(\gamma)}(u,0) du
\]
\[
I_5 \equiv \frac{1}{16\pi B\beta^{2}\rho^{2}}\int_{0}^{\infty}
\frac{1}{\sinh^{2}
[\frac{2\pi}{\beta}\rho\cosh u/2] \cosh^{2}u/2}F_{B}^{(\gamma)}(u,0) du
\]
\[
I_6 \equiv  \frac{1}{32\pi^{2}B\beta\rho^{3}}\int_{0}^{\infty}\frac{ \coth
[ \frac{2\pi}{\beta}\rho\cosh u/2] } { \cosh^{3} u/2 }
F_{B}^{(\gamma)}(u,0) du
\]

For the zero-temperature limit ($\beta\rightarrow \infty$), we obtain

\begin{eqnarray}
< T_{\nu}^{\mu}(x)>_{\beta\rightarrow \infty} & \sim & [\omega_{4}(\gamma) -
\frac{1}{3}\omega_{2}(\gamma)]\frac{1}{\rho^4} diag(1,1,1,-3) \nonumber \\
& & +4(\xi -\frac{1}{6})\omega_{2}(\gamma)\frac{1}{\rho^4}
diag(1,1,-1/2,3/2) ,
\end{eqnarray}
which is the same form presented in ref. \cite{GL}.

In the high-temperature limit ($\beta \rightarrow 0$), we have

\begin{equation}
 < T_{\tau}^{\tau}(x)>_{\beta\rightarrow 0} \sim -\frac{\pi^2}{15\beta^4}
 +(2\xi -1/2)\frac{M^{(\gamma)}}{\beta\rho^3}
\end{equation}
\begin{equation}
 < T_{z}^{z}(x)>_{\beta\rightarrow 0} \sim -\frac{\pi^2}{45\beta^4}
 +\frac{N^{(\gamma)}}{\beta\rho^3}+(2\xi -1/2)
 \frac{M^{(\gamma)}}{\beta\rho^3}
 \end{equation}
\begin{equation}
< T_{\rho}^{ \rho}(x)>_{\beta\rightarrow 0} \sim -\frac{\pi^2}{45\beta^4}+
\frac{N^{(\gamma)}}{\beta\rho^3}-2\xi\frac{ M^{(\gamma)}}{\beta\rho^3}
\end{equation}
\begin{equation}
 < T_{\varphi}^{ \varphi}(x)>_{\beta\rightarrow 0} \sim -
 \frac{\pi^2}{45\beta^4}
-2\frac{N^{(\gamma)}}{\beta\rho^3}+4\xi\frac{M^{(\gamma)}}{\beta\rho^3}
\end{equation}

If we set $\gamma = 0$ in (31-34), our results are in agreement with Frolov
et al. \cite{FR} when $B>1/2$ in their calculations.

\section{Conclusion}

In this paper we have considered the vacuum polarization effect on a
charged scalar field
by a magnetic flux cosmic string at finite temperature.
We
provided the general expression for the
Euclidean thermal heat kernel and its corresponding Green's function
for a massive
scalar field, and a handy form for a massless scalar field. The latter
expression enabled us to evaluate the average of some physical quantities
in a thermal bath.  We calculated the general expression of
the thermal average of $<\phi^{2}(x)>_{\beta}$
and, then, found both zero and high-temperature limits of it.
We have also computed the thermal average of the energy-momentum tensor
of a nonconformal, charged massless scalar field in a general form.
We provided the high-temperature limit to $<T^{\mu}_{\nu}(x)>_{\beta}$.
These formula generalize the work of Frolov et al. \cite{FR} for the case
where there exists an internal magnetic flux to the cosmic string but
it is limited to $B>1/2$.

\section*{Acknowledges}

The author gratefully acknowledges Prof. Bernard Linet for helpful
discussions
and suggestions
and a critical reading of this manuscript. This work was partially
supported by
a grant from CNPq (Brazilian Government Agency).

\end{document}